\documentstyle[aps,psfig]{revtex}

\def\Journal#1#2#3#4{{#1} {\bf #2} (#3) #4}

\def\NPA{{Nucl. Phys. A}}
\def\NPB{{Nucl. Phys. B}}

\def\PRD{{Phys. Rev. D}}

\def\|{\'\i }

\def\kco{k\kern-.4em\raise.4ex\hbox{/}}
\def\kcol{k\kern-.5em\raise.4ex\hbox{/}'}

\title{Unitarized  pion-nucleon scattering amplitude \\ from Inverse Amplitude Method}
\author{Isabela P. Cavalcante\thanks{E-mail: ipc@dfi.ufms.br}}
\address{Universidade Federal de Mato Grosso do Sul, CCET,
Departamento de F\|sica,\\
Cidade Universit\'aria, s/n, Campo Grande, MS, Brazil.} 
\author{J. S\'a Borges\thanks{E-mail: saborges@uerj.br}} 
\address{ Universidade do Estado do Rio de Janeiro, Instituto de
F\|sica, \\
Rua S\~ao Francisco Xavier, 524, Maracan\~a, Rio de Janeiro, RJ, Brazil.}

\begin{document}

\maketitle

\begin{abstract}

In a recent work on low energy pion-nucleon scattering, instead of using chiral perturbation theory (ChPT) amplitude, we started  from  a pion-nucleon {\it 
soft-pion} result  and used  elastic unitarity directly as a dynamical constraint  to construct  first-order unitarity corrected amplitudes.  
The resulting amplitudes are crossing symmetric but, as the ChPT ones, satisfy only approximate unitarity relation. In the present work, we reconsider our approach and we apply the inverse amplitude method (IAM) in order to access the energy resonance region. We present the resulting S- and P-wave phase shifts that are shown to be in qualitative agreement with experimental data. 
\end{abstract}
%%%%%%%%%%%%%%

\section{Introduction}

Even though Quantum Chromodynamics (QCD) is considered nowadays the theory of strong interactions, its application to low energy hadron physics is still far from a solved problem in physics. 
A great theoretical improvement was made by means of the method of Chiral
Perturbation Theory (ChPT) \cite{Leu}, which is an effective theory derived from the basis of QCD. The method consists of writing down chiral Lagrangians for the physical processes and  uses the conventional technique of the field theory for the calculations. 
It is a quite successful method when applied to meson processes. 

In order to deal with baryons in the ChPT approach, the theory faced problems related to fixing the scale for momenta and quark mass expansion. This led eventually to a method known as Heavy Baryon Chiral Perturbation Theory (HBChPT) \cite{ber}, which  has been applied to describe pion-nucleon scattering from a complete effective Lagrangian calculated up to third order in small momenta \cite{Meis}.  
More recently the explicit degrees of freedom corresponding to the
$\Delta$ and $N^*$ resonances have also been considered within HBChPT \cite{dat}. 
One peculiar feature of ChPT is that it leads to partial waves satisfying only approximate elastic unitarity relation, since, for instance, the leading amplitude is a real function of energy in the
physical region. 

 Unitarization methods have been applied to
pion-nucleon scattering in the literature for a long time 
\cite{bade}.  
An interesting approach to implement unitarity in HBChPT amplitude
is the Inverse Amplitude Method (IAM) \cite{dob}. This is a sort of N/D method
that takes the leading contribution to the partial wave amplitude  as    
the numerator $N$ and includes the imaginary part that comes from loop
corrections in the denominator $D$. 
The expansion of $D$ yields the same structure of the original
amplitude up to the order considered, plus corrections of higher
orders, but is fully unitary. 
IAM partial wave amplitudes for pion-nucleon scattering
 have been constructed and fitted to the experimental phase
shifts by fixing nine free parameters \cite{gomez}.  

On the other hand, instead of working in an effective theory
framework,  one can use the hard-meson method of current algebra. 
In fact, current algebra approach was abandoned by several reasons but, as it
is based on chiral symmetry, one expects that it gives equivalent
results for meson processes. In fact, we have compared these two methods and concluded that quasi-unitarized current algebra amplitude is equivalent to ChPT one-loop calculation in the case of kaon-pion scattering and equivalent to one- and two-loop ChPT calculations for pion-pion scattering \cite{bor}. In particular, both methods lead to amplitudes
satisfying the same approximate unitarity relations. 

 Two years ago, we developed an alternative  
procedure to unitarize a current algebra pion-nucleon  scattering
amplitude \cite{bor1}. The unitarity correction was
built as follows. We started from  a {\it
soft-pion} amplitude \cite{osy} 
reproducing  Weinberg  prediction for S-wave scattering
lengths \cite{Wei} and constructed auxiliary functions in order to respect
approximate unitarity.  Considering known the imaginary parts of
partial wave 
amplitudes, we used the dispersion relation technique
 to arrive at a quasi-unitarized amplitude written in terms of known
functions and a {\it two parameter} polynomial part. Furthermore,  we imposed exact crossing
relations to Dirac amplitudes
resulting  very constrained partial waves. 
By this method, even in the absence of free parameters, 
 the violation of unitarity of the resulting amplitudes
seems to be very important mainly in the resonance region.

 The motivation of the present work is to
go beyond threshold by constructing partial wave amplitudes
that respect the unitarity relation exactly. In the
present paper,  we reconsider the unitarity corrected amplitudes described
above in order to
construct IAM modified amplitudes. In the approximation used here, 
each partial wave  requires a maximum of two parameters.

In the next section we present a summary of 
the basic formalism.  In  section III
we present the procedure of Ref. \cite{bor1} for obtaining
quasi-unitary corrected amplitudes. 
In  section IV
we present the IAM unitarization procedure and the results.

\section {Basic formalism}

We consider the reaction \ \ $\pi^a(k)+N(p)\rightarrow
\pi^b(k')+N(p'),$ \ \  
which is described by the  T-matrix amplitude
\begin{equation}
 T^{ab}(p',k';p,k)=-A^{ab}(s,t,u)+{i\over 2}(\kco +\kcol)\
 B^{ab}(s,t,u),
\end{equation}
with $s=(p+k)^2$\ , $t=(k'-k)^2$\ and $u=(p-k')^2$.

In order to specify the various charge states, the invariant
amplitudes $A$\ and $B$\ are decomposed as
\begin{eqnarray}
A^{ab}&=&\delta^{ab}A^{+} +{1\over 2}\lbrack
\tau^{b},\tau^{a}\rbrack A^{-},  \nonumber \\
B^{ab}&=&\delta^{ab}B^{+} +{1\over 2}\lbrack
\tau^{b},\tau^{a}\rbrack B^{-}.  
\end{eqnarray}
They exhibit the following symmetry properties under crossing:
\begin{equation}
 A^{^{\pm}}(s,t,u)=\pm \ A^{^\pm}(u,t,s),\quad 
 B^{^\pm} (s,t,u) = \mp\  B^{^\pm}(u,t,s); 
\end{equation}
the amplitudes corresponding to definite isospin\  $1/2$\
 and $3/2$\ \ are given by
$$ A_{1/2}=A^{+} + 2\,  A^{-}, \ \ \ A_{3/2}= A^{+} - A^{-},$$ 
and similarly for 
\ $B_{1/2} \ , \ B_{3/2}$ . 

We work in the center of mass system, so that the four momenta 
are defined as 
$$ k = (\vec k, w), \ k'= ( \vec {k'},w), \ p
= (-\vec k, E), \ \ {\hbox{and}}\ \ p'= ( \ -\vec{k'}, E),\ 
$$
with  $\vert \vec k \vert = \vert \vec{k'}\vert ,\   w = \sqrt{
\vec k ^2 +m^2}$ \ and \  $E = \sqrt{ \vec k^2 + M^2}$,
\ $M = .938$~GeV  \
and\ $m = .140$~GeV
\ being the nucleon and the pion mass, respectively.
The  total energy and scattering angle are given, respectively,  by \ $ W = E +
w$ \ and\ \ $ \vert \vec k \vert ^2 \cos \theta = \vec k \cdot \vec{k'},$\
thus, in terms of these quantities, one has $ s = W^2$,\  $   t = - 2
\  \vert \vec k \vert ^2 \  ( 1 - \cos \theta)$  \ and\ $u= 2 M^2  + 2
m^2 - s - t $.

For each isospin   $I$  the Pauli amplitudes are
\begin{eqnarray}
F_{{1\, I}} (s,\cos \theta) &=& { {E + M } \over { 8 \pi W }} \bigl
[\ \   A_{I} ( s, \cos \theta) + ( W - M ) B _{I} ( s, \cos \theta)\bigr
], \cr\cr
F_{{2 \, I}} (s,\cos \theta)   &=&  { {E - M }
\over { 8 \pi W }} \bigl [- A_{I} ( s, \cos \theta) + ( W + M )
B_{I} ( s, \cos \theta)\bigr ] .
\end{eqnarray}
Partial wave amplitudes $f^{^\pm}_{_{I\,\ell}} $ \ are defined as 
\begin{eqnarray}
 f^{\pm}_{{I\, \ell}}(s)& =& F_{{1\, I\, \ell}}(s) +
F_{{2 \, I\, \ell\pm 1}}(s), \quad \mbox{where} \cr
 F_{{i\, I\,  \ell}} (s)& =& {1\over 2} \int_{-
1}^{+ 1} F_{{i\, I}} (s,x) P_\ell (x)\  dx, \ \ {\hbox{for}}
\ i = 1,2 \ {\hbox{and}} \ I= \frac 1 2 ,  \frac 3 2 .
 \label{def-pw}
\end{eqnarray}

 For elastic scattering we have 
$\mbox{Im} \, f_{{I\, \ell}}^{\pm} (s) = \vert \vec k \vert \ \vert
f_{{I\,\ell}}^{\pm} (s) \vert ^2,$ %ref unit  
which may be solved yielding
\begin{equation} f_{{I\, \ell}}^{\pm} (s) = {1 \over {\vert \vec k
\vert }}\  e\, ^ { i \delta_{I\,\ell}(s)} \ \sin \delta_{I\, \ell}
(s),
\end{equation}
where $\delta_{I\, \ell}(s)$ are real phase shifts. 

\section{Unitarization procedure}

Let us recall the main points in our unitarization procedure. First,
we consider an amplitude reproducing S-wave scattering lengths
predicted by Weinberg. 
The low energy amplitude we  start from 
is the  {\it soft-pion} limit 
obtained  by Osypowsky \cite{osy} using  the Ward identity
technique. The final expression for the four point
function,  related to
 pion-nucleon scattering,  is written in terms of form factors and
 propagators. By estimating the contribution of each  term at
threshold one is led to assume that the low energy
 amplitudes are
\begin{equation}
 A^{{ca\, - }} = {\mu_{_V}\over {8 M f^2}} (u - s), 
\qquad B^{{ca\, -}} = { 1+\mu_{_V}\over {2 f^2}}, \qquad A^{ca\, +} \ = \ B^{ca\, +}\ = \ 0, 
\end{equation}
where $f = .094$~GeV  is the pion decay constant,  $\mu_{_V}
\simeq 3.7$ and the superscript {\it ca} stands for {\it current algebra}. 
 These amplitudes lead to the well known Weinberg
prediction for S-wave scattering lengths, namely
\ \ $ a_0^{+} = 0$ \ \ and\ \ $ a_0^{-} = 0.077 \ m^{-1}$, to be
compared with the experimental values \ \ $a_0^+ = -0.015 \pm 0.015
\ m^{-1} $\ and \ $a_0^- = 0.097 \pm 0.003 \ m^{-1}$.

 We implemented unitarity for the low-energy pion-nucleon
amplitude  $A^{ca\, -}$\ and $B^{ca\, -}$. From our previous
analysis on meson scattering, we conjecture that the corrected
amplitudes must  satisfy 
$$A_\ell (s) \simeq A^{{ca}}_\ell (s)+
A_\ell^{{(1)}}(s) + {\cal O}(\epsilon ^2),\quad {\hbox{for}}\quad  s\simeq (
m_\pi + m_N)^2, \quad {\hbox{ and the same for}}\ B_\ell; $$
$A^{{(1)}}$\ is a complex function and $\epsilon$\ is a small
parameter characterizing the corrections.

For each isospin channel, {\it soft-pion} amplitudes are obtained from (\ref{def-pw}) as
\begin{eqnarray*}
f_{_0}^{ ca\, +}(s) &=& F_{_{1\, 0}}^{ca} (s) + F_{_{2\, 1}}^{ca}
(s), \quad f_{_2}^{ ca\, -} (s) = F_{_{2\, 1}}^{ca} (s), \\
f_{_1}^{ ca\, -} (s) &=& F_{_{1\, 1}} ^{ca}(s) + F_{_{2\, 0}}^{ca} (s),
\quad f_{_1}^{ ca\, +} (s) = F_{_{1\, 1}}^{ca}(s),
\end{eqnarray*}
where \quad $f_\ell^{ca\ \pm}$\quad follow from \ $A^{ca}$\ and \ $B^{ca}$.
It is evident that \ Im~$f_\ell^{ca\, \pm} (s) = 0$, but this situation 
is inherent to {\it soft-meson} calculation.

We construct quasi-unitarized amplitudes by 
requiring that the corrections \ $F_{i\, \ell}
^{(1)}$,  for \ $s\ge (M+m)^2$, have the following imaginary parts:
\begin{eqnarray*}
{\hbox{Im}}\, F_{{1\, 0}}^{(1)} (s) & = & \vert \vec k \vert \bigl(
F_{{1\,0}}^{{ ca\, 2}}(s) + 2 F_{{1\, 0}}^{{ca}}(s) F_{{2\,
1}}^{{ca}}(s)\bigr), \\ {\hbox{Im}}\,  F_{{2\, 0}}^{{(1)}} (s) &
= &  \vert \vec k \vert \bigl ( F_{{2\, 0}}^{{ ca\, 2}}(s) + 2\,
F_{{2\, 0}}^{{ca}} F_{{1\, 1}}^{{ca}}(s)\bigr),\\ 
{\hbox{Im}}\,
F_{{1\, 1}}^{{(1)}} (s) &  = & \vert \vec k \vert F_{{1\, 1}}^{{
ca\, 2}}(s), \qquad \quad  {\hbox{Im}}\
F_{{2\,1}}^{{(1)}} (s) = \vert \vec k \vert F_{{2\, 1}}^{{ ca\,
2}}(s).
\end{eqnarray*}
It guarantees that 
the quasi-unitarized partial waves \
$f_\ell^{\pm} = f_\ell^{ca\, \pm} + f_\ell^{(1)\, \pm}$\
satisfy 
\begin{equation}
 {\hbox{Im}}\ f_\ell^{\pm}
%\, (1)} 
= {\vert \vec k \vert } \
\vert f_\ell^{\pm\, ca}\vert
^2.
\end{equation}

Expressing functions $A$\ and $B$\ in terms of Pauli amplitudes, we
construct the auxiliary functions
\begin{eqnarray}
 {\cal  A } (s,\cos \theta_{s})  =
{1\over 4}\bigl [ a_1(s) {\cal  S}(s) + a_2(s){\cal  D} (s) + 3 \cos
\theta_{s} a_3(s) {\cal Q}(s)\bigr],\cr
{ \cal
B}(s,\cos \theta_{s})  =   {1\over 4}\bigl[ b_1(s) {\cal  S} (s) +
b_2(s) {\cal  D} (s) + 3 \cos \theta_{s} b_3(s) {\cal Q}
(s)\bigr],
\end{eqnarray}
where 
$${\hbox{Im}}\  {\cal S} (s)  = {{2 \vert \vec k \vert}\over W}
\ A_0^{ca}(s), \quad {\hbox{Im}}\  {\cal  D }  (s)  = {{2 \vert \vec k \vert}\over W}
\ B_0^{ca}, \  \  {\cal Q } (s) = {{2 \vert \vec k
\vert}\over W}A_1^{ca} (s),$$
%%% 
\begin{eqnarray*}
 a_1(s) & =  & (W+M) \bigl(\, F_{{1\, 0}}^{{ca}} + 2 F_{{2\,
1}}^{{ca}}\bigr ) \, + \, ( W - M ) ( F_{{2\, 0}}^{{ca}}+ 2
F_{{1\, 1}}^{{ca}}\bigr), \\ a_2 (s) & = & (W^2 - M^2) \bigl( \,
F_{{1\, 0}}^{{ca}} + 2 F_{{2\, 1}}^{{ca}} - F_{{2\, 0}}^{{ca}}
- 2 F_{{1\, 1}}^{{ca}}\bigr),\\ a_3(s) & =  & (W + M) F_{{1\,
1}}^{{ca}} + (W - M ) F_{{2\, 1}}^{{ca}}, \\ b_1(s) &  =  &
F_{{1\, 0}}^{{ca}} + 2 F_{{2\, 1}}^{{ca}} - F_{{2\, 0}}^{{ca}}
- 2 F_{{1\, 1}}^{{ca}}, \\ b_2 (s) & = & ( W - M) \bigl( F_{{1\,
0}}^{{ca}} + 2 F_{{2\, 1}}^{{ca}}\bigr) \ +\ (W + M) \bigl(
F_{{2\, 0}}^{{ca}}+ 2 F_{{1\, 1}}^{{ca}}\bigr),\\ b_3(s) & = &
F_{{1\, 1}}^{{ca}} - F_{{2\, 1}}^{{ca}}.
\end{eqnarray*}

In order to avoid kinematical singularities, we write subtracted
dispersion relations for ${\cal  S}$, ${\cal D}$ and ${\cal Q}$ by
introducing free parameters.
At this point the present procedure must be handled in a
different manner than done in Ref. \cite{bor1}. In the construction of 
the quasi-unitarized amplitudes  we introduced only two parameters. On the other hand,  in the
present work, one must use up to two parameters for {\it each} partial wave, chosen from the set of parameters
($\lambda_1, \lambda_2, \lambda_3$) in the expressions 
$$ {\cal  S}(s) = s^2 \lambda_1 +
A_0^{ca}(s)\ G(s),\quad {\cal  D}(s) = s^2 \lambda_2 + B_0^{ca}\ G(s), \quad
 {\cal Q}(s) = s^2 \lambda_3  + A_1^{ca} (s)\ G(s),$$
with
$$ G(s) = {s^3\over \pi} \int_{(M+m)^2}^\infty \
{\sqrt{\bigl[x-(M+m)^2\bigr] \bigl[x-(M-m)^2\bigr]}\over{x^4\ (x - s
)}} \ dx.$$
One may argue that in the definition of the
subtracted dispersion relations above, 
terms in powers of $s$ smaller than two are missing. 
This restricted choice of the number of 
subtraction constants is based on the fact that in the fitting process
those extra free parameters do not provide qualitative improvement on the
fits.

As done in our previous work, 
crossing properties and partial wave total isospin
dependence are imposed to the corrected amplitudes, by
taking 
\begin{eqnarray*}
 A^{(1)\, +} (s,t,u )&  = & 2 \,  {\cal{A}}(s, t) +
(s \leftrightarrow u  ),\ \ A^{(1)\, - } (s,t,u ) = {\cal{A}}(s, t) -
(s \leftrightarrow u  ), \\
 B^{(1)\, +} (s,t,u ) & =
& 2\,  {\cal{B}}(s, t ) - (s \leftrightarrow u ),\ \
B^{(1)\, - } (s,t,u ) =  {\cal{B}}(s, t)  +  (s
\leftrightarrow u  ).
\end{eqnarray*}
The corrections to partial wave amplitudes are calculated
from (5) using (4) and are indicated by $f^{(1)}_{\ell\, I} (s)$.
This was the final step of the procedure introduced in Ref. \cite{bor1}. 

\section{Inverse Amplitude Method and results}

The quasi-unitarized amplitudes from last section do not obey 
{\it exact} unitarity relation.
In order to restore it, we apply the IAM to
the corrected partial waves $f_{\ell \, I} = f_{\ell \, I}^{ca} +
f_{\ell \, I}^{(1)}$,  by writing
$$\tilde f_{\ell \, I} (s) = \frac {f_{\ell \,I}^{ca}(s)}{ 1 -  
f_{\ell \,I}^{(1)} (s) / f_{\ell \,I}^{ca}(s)}.$$

In order to fit IAM amplitudes to the experimental results of phase shifts one has to be cautious. As the inelasticity for some partial waves becomes important for energies less than 1.5 GeV, one has to introduce some criteria to define the validity region of the IAM. In the present paper we decided to fit our results to experimental phase shifts corresponding to dimensionless experimental amplitudes $f_{\ell \, I}^{exp}$ satisfying $ {\hbox{Im}}\  f_{\ell \, I}^{exp} - \vert  f_{\ell \, I}^{exp} \vert ^2 < 0.1$.

Restricting ourselves to these ranges in energy, we performed the fits of 
 $S_{1\,1}, S_{3\, 1},
P_{1\, 1}, P_{3\, 1}, P_{1\, 3}  \ {\hbox{and}}\
 P_{3\, 3}$  partial waves from the form 
$\tilde f_{\ell \, I}$ above to experimental data, thus fixing the
 free parameters $\lambda_1$,  $\lambda_2$  and $\lambda_3$
 independently, but only a maximum of {\em two} of them  for each wave. Their values are presented in the Table. 

\begin{table}
\begin{center}
\begin{tabular}{lrrrrrr}
  &   S11   &   P11  &  P13  &  S31  &  P31  & P33  \\ 
            \hline
$\lambda_1$ \hspace{.5cm} & $-15.5 $ & $-$  & $-$ & $-$ & $-$ &   $-3.8$   \\
$\lambda_2$ \hspace{.5cm} & $24.3$ &$ 39.8$ & $-$ & $2.3$ & $-48.0$ & $-$  \\ 
$\lambda_3$ \hspace{.5cm} & $-$ & $16.3$  & $-6.9 $ & $-8.5$ & $-27.0$ & $-8.3$
\end{tabular}
\end{center}
\caption{Results from fits of IAM modified quasi-unitarized partial
waves to experimental data; $\lambda_1$ and $\lambda_3$ are given in
units of (GeV)$^{-5}$, while $\lambda_2$, in units of (GeV)$^{-6}$.} 
\end{table}

The phase shifts thus obtained are shown in the figures, as functions
of the cms energy, in GeV. In the same figures we plot the phase
shifts obtained with the original quasi-unitarized amplitudes \cite{bor1}. In that case only $\lambda_1$ and  $\lambda_3$ were considered, and their values were the same for all partial waves. 
Concerning our present results, we observe that it is possible to fit the  phase
 shifts of our model to the experimental data (Ref. \cite{exp}).
We also observe that,  using the IAM,  it is not possible  to simultaneously fit two waves with the same values of parameters.
The qualitative agreement with experimental data is an indication that
the IAM allows one to access the resonance region of pion-nucleon scattering 
since  some low energy chiral amplitude is given.

We would like to emphasize that in the construction of the quasi-unitarized amplitude  there is no commitment on the
existence of any resonance. They emerge as dynamical
consequences of the unitarization procedure. Notice also that we do
not include the nucleon pole in the low energy amplitude we started
from. 

 In order to further explore the results that we have
obtained, one should construct corrections in the next level of 
unitarity approximation for the
partial waves, as outlined in the last equation of Ref. \cite{bor1}.
We believe that the IAM applied to this new result will allow one to
access the energy resonance region for higher angular momenta partial
waves.

\newpage
\begin{figure}[htb]
\centerline{\psfig{figure=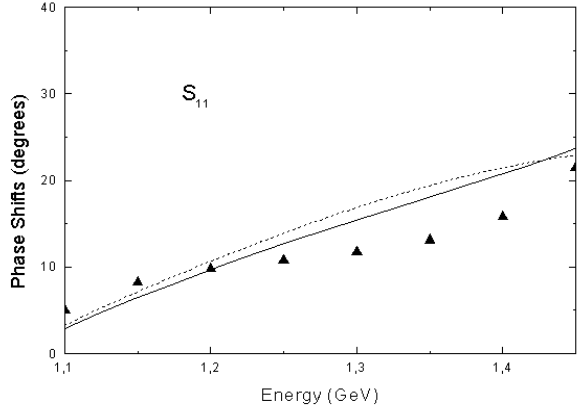,width=8.9cm}\psfig{figure=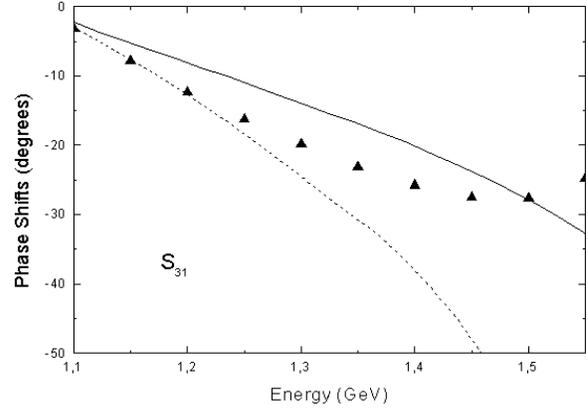,width=8.9cm}}
\caption{S-wave phase shifts (in degrees) as functions of cms energy
(in GeV); present results (solid) and previous quasi-unitary ones (dashed). }
\end{figure}

\begin{figure}[htb]
\centerline{\psfig{figure=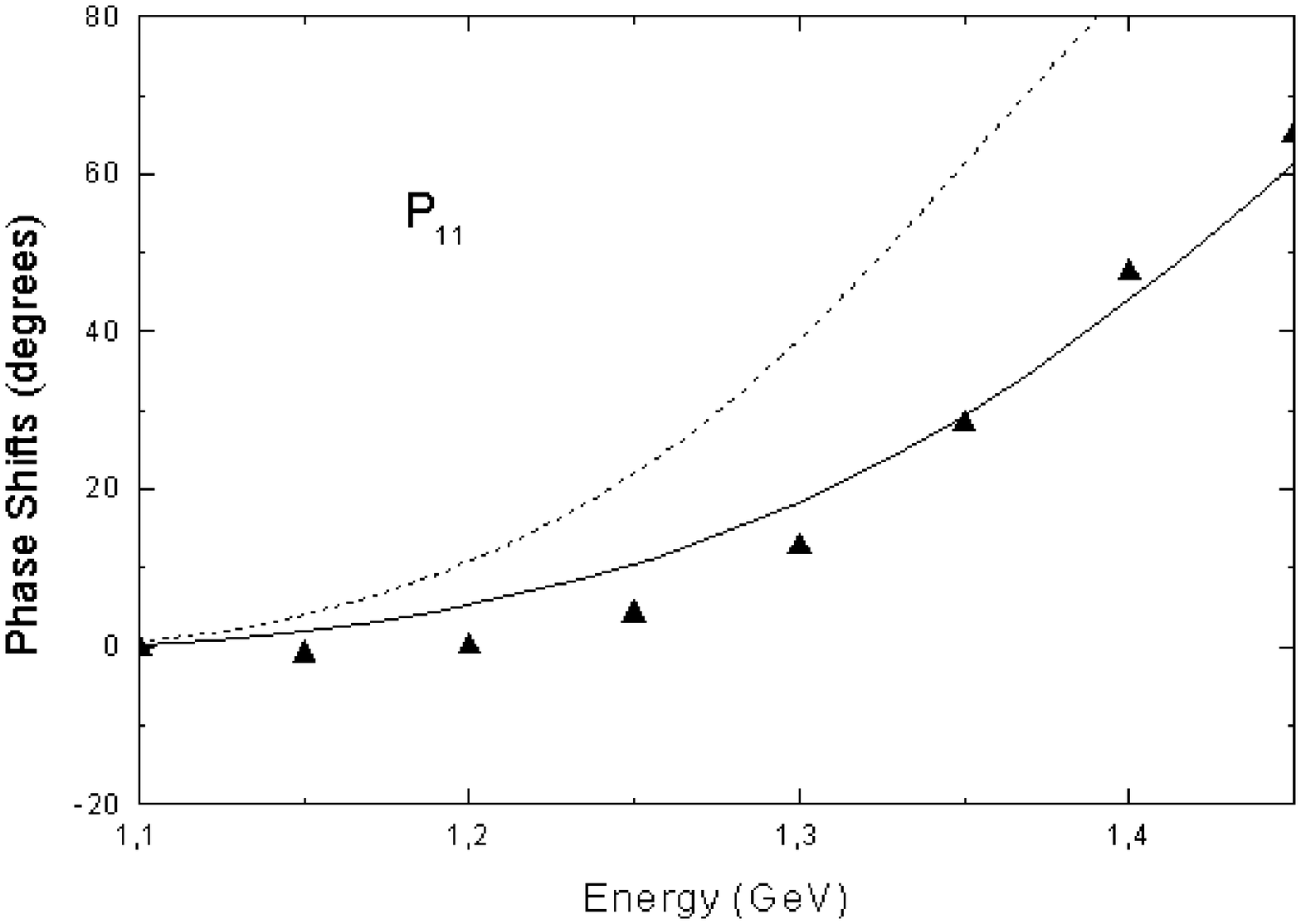,width=8.9cm}\psfig{figure=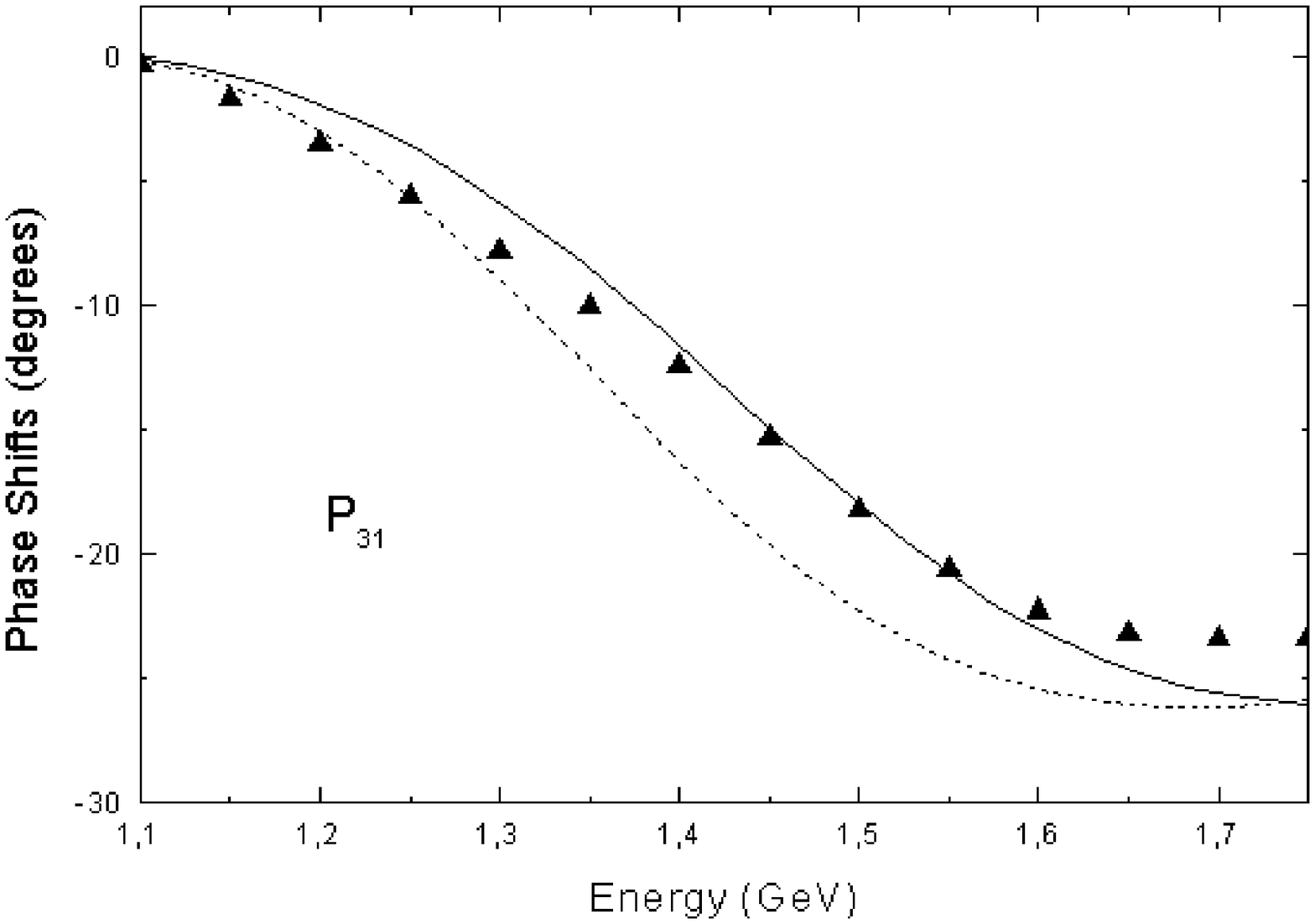,width=8.9cm}}
\caption{Isospin $1/2$ P-wave phase shifts (in degrees) as functions of cms energy
(in GeV); present results (solid) and previous quasi-unitary ones (dashed).}
\end{figure}

\begin{figure}[htb]
\centerline{\psfig{figure=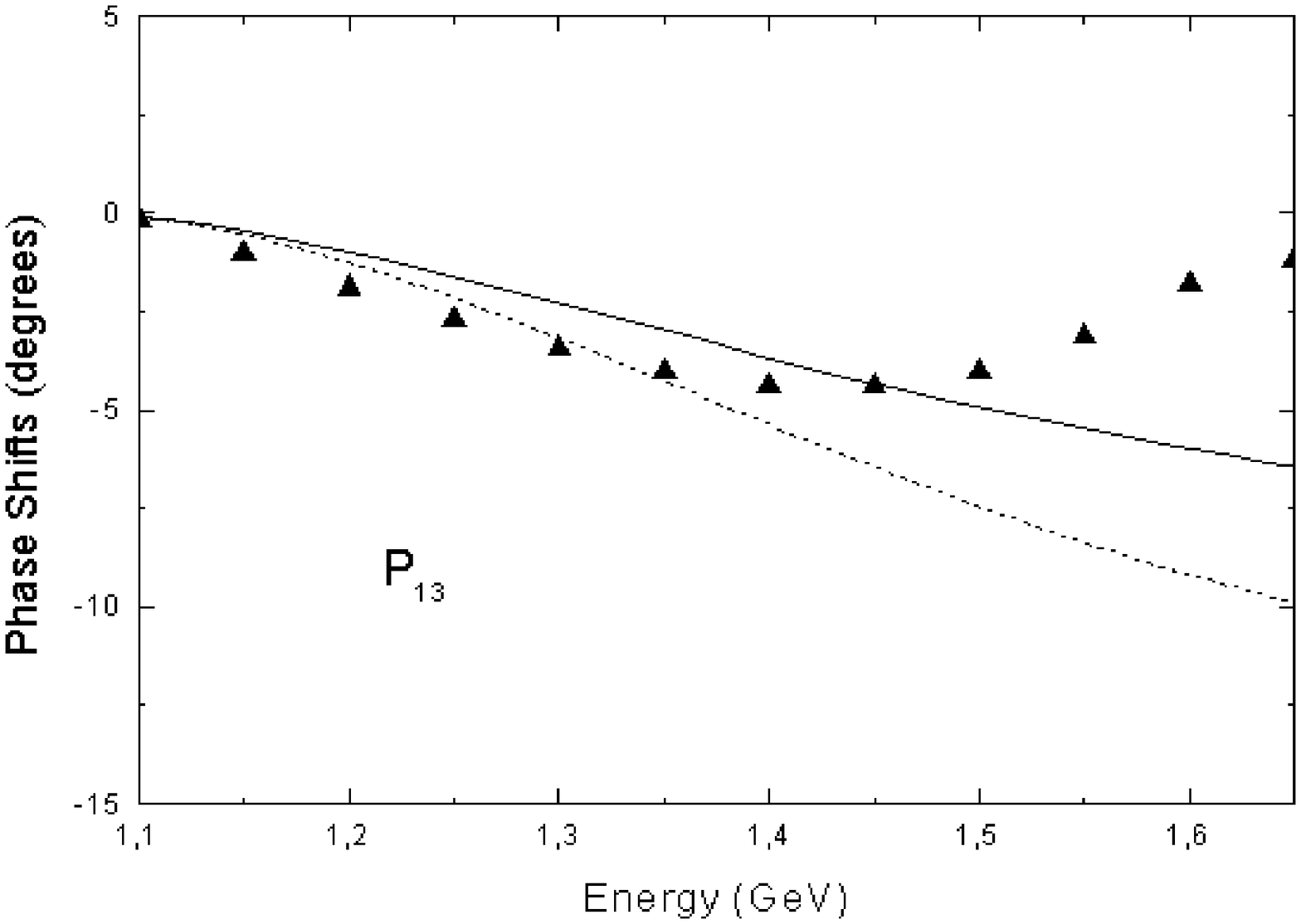,width=8.9cm}\psfig{figure=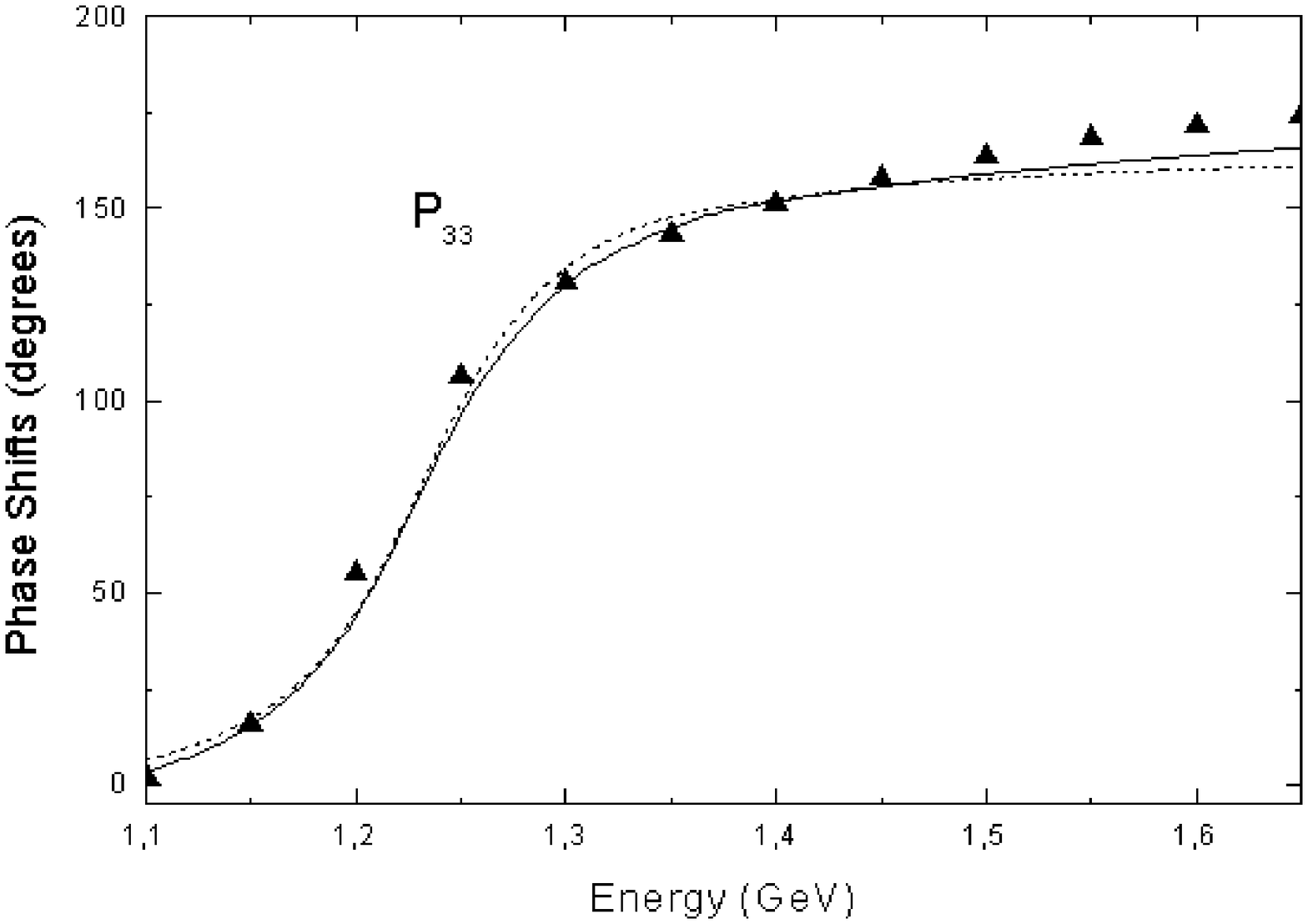,width=8.9cm}}
\caption{Isospin $3/2$ P-wave phase shifts (in degrees) as functions of cms energy
(in GeV); present results (solid) and previous quasi-unitary ones (dashed).}
\end{figure}

\end{document}